\documentclass[preprint,aps,nofootinbib,prd,showpacs]{revtex4}
\usepackage{graphicx}
\usepackage{amsmath}
\usepackage{amsfonts}
\usepackage{amssymb}
\usepackage{color}%
\usepackage{slashed}

\providecommand{\U}[1]{\protect\rule{.1in}{.1in}}

\newcommand{\f}{\begin{equation}}
\newcommand{\ff}{\end{equation}}
\newcommand{\fa}{\begin{eqnarray}}
\newcommand{\ffa}{\end{eqnarray}}

\begin{document}
\title{Anomaly freedom of the vector modes with holonomy corrections in perturbative Euclidean loop quantum gravity}
\author{Jian-Pin Wu $^{1}$}
\email{jianpinwu@bnu.edu.cn}

\author{Yongge Ma $^{1}$}
\email{mayg@bnu.edu.cn}
\affiliation{$^1$Department of Physics, Beijing Normal University, Beijing 100875, China}

\begin{abstract}

We study the perturbation of the effective Hamiltonian constraint with holonomy correction from Euclidean loop quantum gravity. The Poisson bracket between the corrected Hamiltonian constraint and the diffeomorphism constraint is derived for vector modes. Some specific form of the holonomy correction function $f^{i}_{cd}$ is found, which satisfies that the constraint algebra is anomaly free. This result confirms the possibility of nontrivial holonomy corrections from full theory while preserving anomaly-free constraint algebra in the perturbation framework. It also gives valuable hints on the possible form of holonomy corrections in the effective loop quantum gravity.

\pacs{04.60.Pp, 04.60.Kz, 98.80.Qc}

\end{abstract}

\maketitle

\section{Introduction}

It is well known that general relativity (GR) is a totally constrained system with first-class constraints.
In connection dynamical formalism, the constraint algebra of GR takes the form
\begin{equation} \label{CAClassicalLevel}
\{\mathcal{C}_{I},\mathcal{C}_{J}\}=\mathcal{K}^{K}_{IJ}(A^{j}_{b},E^{a}_{i})\mathcal{C}_{K} ~,
\end{equation}
where $\mathcal{C}_{I}$ are the smeared constraints (Gauss constraint, diffeomorphism constraint and Hamiltonian constraint), and $\mathcal{K}^{K}_{IJ}(A^{j}_{b},E^{a}_{i})$ are, in general, structure functions of the phase space variables $(A^{j}_{b},E^{a}_{i})$. In order to have a well-defined physical behavior, the algebra should also be closed at the quantum level. In the canonical approach to quantize GR, such as loop quantum gravity (LQG), one would expect to represent the above constraint algebra on some kinematical Hilbert space.

As a nonperturbative and background-independent quantum gravity theory \cite{LQGRovelli,LQGThiemann,LQGAshtekar,LQGHan}, LQG has received increased attention recently. In
the symmetry-reduced models of LQG, known as LQC\cite{LQC1,LQC2,LQC3,LQC4,LQC5,LQC6}, the study of effective theories has become topical since it may relate the quantum gravity effects to low-energy physics. The effective equations of LQC are being studied from both the canonical perspective\cite{Taveras,DMY,YDM,BT09,BT10,Boj11} and the path integral perspective\cite{ACH09,ACH101,ACH102,QHM,QDM,QM1,QM2}.
In general, two main quantum gravity effects, namely
the inverse volume correction and the holonomy correction, would appear in the effective Hamiltonian constraint of LQC. Due to the introduction of quantum effects, the corresponding constraint algebra
might not close but has a so-called anomaly term, $\mathcal{A}_{IJ}$,
\begin{equation} \label{CAQuantumLevel}
\{\mathcal{C}_{I},\mathcal{C}_{J}\}=\mathcal{K}^{K}_{IJ}(A^{j}_{b},E^{a}_{i})\mathcal{C}_{K} + \mathcal{A}_{IJ} ~.
\end{equation}
As pointed out in Ref.\cite{AnomalyFreedom}, the anomaly would obstruct the purpose of cosmological perturbation theory based on effective LQG, since the quantum corrected perturbation equations could not be expressed solely in terms of gauge-invariant variables.

On the other hand, to have a good understanding for the structure formation and anisotropies of the cosmic microwave background (CMB), one needs to consider the linear perturbations around Friedmann-Robertson-Walker (FRW) spacetimes. Therefore, it is very interesting and valuable to obtain an anomaly-free constraint algebra of cosmological perturbations with loop quantum effects. For inverse volume correction of LQC, the anomaly-free constraint algebra and the corresponding gauge-invariant cosmological perturbation equations have been derived for scalar modes \cite{AnomalyFreedom,GaugeInvariant}, vector modes \cite{VectorBojowald} and tensor modes\footnote{In fact, for the tensor modes, the anomaly-free constraint algebra is automatically fulfilled.} \cite{TensorBojowald}, respectively. Along this direction, it is worthwhile to point out that some relevant applications, including the primordial power spectrum and non-Gaussian, have already been investigated intensively \cite{ObservablesBojowald,ObservablesCBojowaldPRL,ObservationalTBojowald,NonGaussianLQC}.

For the holonomy correction, some pioneer works have been done to study the anomaly-free constraint algebra.
For vector modes, in Ref.\cite{VectorBojowald}, it was shown that an anomaly-free algebra is satisfied up to the fourth order of the background extrinsic curvature variable $\bar{k}$. However, it becomes less reliable for vector modes to propagate through the cosmic bounce.
Also, an anomaly-free constraint algebra is obtained when the higher-order holonomy is included\cite{HolonomyVectorH}.
In addition, by introducing the counterterms in the Hamiltonian constraint,
it was shown that the anomaly-free constraint algebra can also be satisfied\cite{HolonomyVectorA}.
For scalar modes, the situation becomes more complicated. However, scalar modes are more interesting and valuable because they could be related to some observables, such as the power spectrum and the non-Gaussianity.
A tentative attempt was made in Ref.\cite{HolonomyJPWu} to derive the cosmological
perturbation equations for scalar modes with holonomy corrections in longitudinal gauge. The result shows that the holonomy effects influence both background and perturbations and contribute
the nontrivial quantum corrected terms in the cosmological perturbation equations.
However, in order to
obtain the consistent cosmological perturbation equations, one need add some special and
nonunique terms to them.
In order to cure the shortcoming of previous works, an effective Hamiltonian with a new holonomy correction was introduced in Ref\cite{HolonomyScalarEwing}, where an anomaly-free constraint algebra is obtained. But the method was performed in the longitudinal gauge, and the extension to the gauge-invariant case is not straightforward.
In Ref.\cite{HolonomyScalarA}, by using the same method developed in Ref.\cite{HolonomyVectorA} for vector modes, i.e., adding the counterterms, an anomaly-free constraint algebra for scalar modes with holonomy corrections was obtained and the gauge-invariant cosmological perturbation equations were derived.
However, it should be noted that in all above-mentioned works,
the so-called holonomy corrections are only included after, rather than before, doing perturbations of the classical Hamiltonian constraint. Thus the resulting anomaly-free cosmological perturbation theory would only contain partial holonomy corrections, though it could give certain hints to a full treatment.
In contrast, in this paper, we will first propose an effective Hamiltonian constraint with holonomy corrections from full LQG and then perturb it directly to obtain the cosmological perturbation equations. So our aim is to obtain the complete cosmological perturbation theory with holonomy corrections from full LQG.
But it should also be noted that it is difficult to derive an effective Hamiltonian from full LQG. As a first step, we will consider only the possible holonomy corrections in Euclidean LQG. The Lorentzian case is left for future study.  Also, we will first focus on the vector modes, while the scalar modes will be addressed elsewhere. Some specific form of the holonomy correction function will be proposed, which satisfies that the perturbative constraint algebra is anomaly free.

\section{The correction function of full theory}

The connection dynamical formalism of GR is subject to the Gaussian, diffeomorphism and Hamiltonian constraints\cite{LQGAshtekar,LQGHan}. Since the Gaussian constraint forms an ideal in the constraint algebra, in the kinematical treatment of LQG one may easily work in the internal gauge-invariant Hilbert space where the Gaussian constraint has been implemented. Moreover, since there is no diffeomorphism constraint operator in the kinematical Hilbert space, one usually considers finite diffeomorphism transformations instead of the diffeomorphism constraint to construct diffeomorphism invariant states by the group-averaging procedure. Based on the above treatment in LQG, it is reasonable to first consider only the holonomy correction in the Hamiltonian constraint.

In the canonical formulation, the gravitational Hamiltonian constraint can be written as
\begin{eqnarray}
\label{GravityHamiltonian}
H_{G}[N]=\frac{1}{16\pi G}\int _{\Sigma}d^{3}x N
\epsilon^{jk}{}_{i}\frac{E^{c}_{j}E^{d}_{k}}{\sqrt{|det E|}}\left[F^{i}_{cd}
-(\gamma^{2}-s)\epsilon^{i}{}_{mn}K^{m}_{c}K^{n}_{d}\right],
\end{eqnarray}
where the curvature of the Ashtekar-Barbera connection is given by
\begin{eqnarray} \label{Curvature}
F^{i}_{cd}=2\partial_{[c}A^{i}_{d]}
+\epsilon^{i}{}_{mn}A^{m}_{c}A^{n}_{d}.
\end{eqnarray}
In Euclidean GR, the signature $s=1$ and the simplest selection of the Barbero-Immirzi parameter is
$\gamma=\pm 1$ (we will adopt $\gamma=1$ for convenience).
Then the Hamiltonian density becomes
\begin{eqnarray}
\label{GravityHamiltonianD}
\mathcal{H}_{EG}=\epsilon^{jk}{}_{i}\frac{E^{c}_{j}E^{d}_{k}}{\sqrt{|det E|}}F^{i}_{cd}.
\end{eqnarray}
In LQG, the fundamental variables are holonomies and triad fluxes. Thus the connection $A^{i}_{a}$ would be replaced by the corresponding holonomy,
\begin{eqnarray}
\label{Holonomy}
h_{e}(A)=\mathcal{P}\exp\int_{e}A^{i}_{a}\tau_{i}dx^{a},
\end{eqnarray}
where the symbol $\mathcal{P}$ denotes path ordering, and $\tau_{j}=-\frac{i}{2}\sigma_{j}$ is a basis in
the algebra $su(2)$ with $\sigma_{j}$ being the Pauli matrices. Correspondingly, the effective curvature $F^{i}_{cd}$ would be modified by the holonomy corrections. Therefore,
we could consider in general the following effective holonomy corrections to the Euclidean Hamiltonian
\begin{eqnarray}
\label{EuclideanHdwithHolonomy}
\mathcal{H}^{Q}_{EG}=\epsilon^{jk}{}_{i}\frac{E^{c}_{j}E^{d}_{k}}{\sqrt{|det E|}}f^{i}_{cd}
(A,\partial A,\partial^{2} A,\cdots,\partial^{n} A,\epsilon),
\end{eqnarray}
where $f^{i}_{cd}(A,\partial A,\partial^{2} A,\cdots,\partial^{n} A,\epsilon)\equiv F^{i}_{cd}(h_{e}(A))-F^{i}_{cd}(A)$
is an arbitrary function of $A^{m}_{a}$ and its derivatives.
In addition, it is natural to assume that the holonomy-correction function $f^{i}_{cd}(A^{m}_{a},\epsilon)$
is also an antisymmetrical tensor as $F^{i}_{cd}$ is. So, the corrected Hamiltonian constraint can be reexpressed as
\begin{eqnarray}
\label{EuclideanHhC}
H^{Q}_{EGT}[N]=\frac{1}{16\pi G}\int _{\Sigma}d^{3}xN[\mathcal{H}_{EG}+\mathcal{H}_{hEG}]:=H_{EG}[N]+H^{Q}_{EG}[N].
\end{eqnarray}

\section{The perturbative Euclidean loop quantum gravity}

\subsection{The basic variables}

In loop quantum gravity, instead of the spatial metric $q_{ab}$, a
densitized triad $E^{a}_{i}$ is primarily used, which satisfies
$E^{a}_{i}E^{b}_{i}=q^{ab}detq$. Moreover, in the canonical
formulation the spacetime metric is given by
\begin{equation} \label{CSTM}
ds^{2}=-N^{2}d\eta^{2}+q_{ab}(dx^{a}+N^{a}d\eta)(dx^{b}+N^{b}d\eta) ~,
\end{equation}
where $N$ and $N^{a}$ are lapse function and shift vector
respectively.
By comparing the above equation with the spatially flat FRW metric
\begin{equation} \label{IHFRWM}
ds^{2}=a^{2}(\eta)(-d\eta^{2}+\delta_{ab}dx^{a}dx^{b}) ~,
\end{equation}
the background variables, $\bar{N}$, $\bar{N^{a}}$ and
$\bar{E}^{a}_{i}$, can be, respectively, expressed as
\begin{equation} \label{BV1}
\bar{N}=\sqrt{\bar{p}};~~ \bar{N^{a}}=0;~~ \bar{E}^{a}_{i}=\bar{p}\delta^{a}_{i} ~~,
\end{equation}
where the background variables are denoted with a bar, which
describes smoothed out, spatially averaged quantities. Another
background variable, the extrinsic curvature components
$\bar{K}^{i}_{a}$, can be expressed by
\begin{equation} \label{EC1}
\bar{K}_{ab}=\frac{1}{2\bar{N}}(\dot{\bar{q}}_{ab}-2D_{(a}\bar{N}_{b)})=\dot{a}\delta_{ab},
\end{equation}
where $D$ is the covariant spatial derivation. Thus, one has
\begin{equation} \label{BV2}
\bar{K}^{i}_{a}
=\frac{\bar{E}^{b}_{i}}{\sqrt{|det(\bar{E}^{c}_{j})|}}\bar{K}_{ab}
=\frac{\dot{\bar{p}}}{2\bar{p}}\delta^{i}_{a}
=:\bar{q}\delta^{i}_{a} .
\end{equation}
In Eq.(\ref{BV2}), we have defined the background extrinsic
curvature as
$\bar{q}=:\frac{\dot{\bar{p}}}{2\bar{p}}=\frac{\dot{a}}{a}$, which can
also be obtained from the background equations of motion
\cite{GaugeInvariant}.
At the same time, from the full expression of the spin connection
\begin{equation} \label{Spinconnection}
\Gamma^{i}_{a} = -\frac{1}{2} \epsilon^{ijk} E^{b}_{j}
\left( 2\partial_{[a} E^{k}_{b]} + E^{c}_{k} E^{l}_{a} \partial_{c} E^{l}_{b} - E^{k}_{a} \frac{\partial_{b}(det E)}{det E}\right)~,
\end{equation}
we can conclude that the background variable $\bar{\Gamma}^{i}_{a}$ vanishes.
Therefore, the background connection variables $\bar{A}^{i}_{a}$ can be diagonal, and hence the full connection
can be expanded as
\begin{equation} \label{BPcurvature1}
A^{i}_{a}=\bar{A}^{i}_{a}+\delta A^{i}_{a}=\bar{q}\delta^{i}_{a}+\delta A^{i}_{a}.
\end{equation}
Similarly, the densitized triad $E^{a}_{i}$ can also be expanded as
\begin{equation} \label{BPEaiExpansion}
E^{a}_{i}=\bar{E}^{a}_{i}+\delta E^{a}_{i}=\bar{p}\delta^{a}_{i}+\delta E^{a}_{i}.
\end{equation}
In addition, the homogeneous mode is defined by
\begin{equation} \label{Hmode}
\bar{p}:=\frac{1}{3V_{0}}\int_{\Sigma} E^{a}_{i}\delta^{i}_{a}d^{3}x,~~~ \bar{q}:=\frac{1}{3V_{0}}\int_{\Sigma} A^{i}_{a}\delta^{a}_{i}d^{3}x~,
\end{equation}
where we integrate over a bounded region of coordinate size
$V_{0}=\int_{\Sigma} d^{3}x$.
Then by using the above Eqs.(\ref{BPcurvature1}), (\ref{BPEaiExpansion}) and (\ref{Hmode}), we will find that
$\delta E^{a}_{i}$ and $\delta A^{i}_{a}$ do not have homogeneous modes, namely
\begin{equation} \label{PHmode}
\int_{\Sigma} \delta E^{a}_{i}\delta^{i}_{a}d^{3}x=0,~~~ \int_{\Sigma} \delta A^{i}_{a}\delta^{a}_{i}d^{3}x=0,
\end{equation}
Therefore, we can construct the Poisson brackets of the background and perturbed variables as
\cite{AnomalyFreedom}
\begin{equation} \label{PB}
\{\bar{q},\bar{p}\}=\frac{8\pi G}{3V_{0}},~~~ \{\delta A^{i}_{a}(x),\delta E^{b}_{j}(y)\}=8\pi G\delta^{i}_{j}\delta^{b}_{a}\delta^{3}(x-y)~.
\end{equation}
In addition, we would like to point out that for a similar reason, the perturbed lapse $\delta N$ do not
have homogeneous modes either,
\begin{equation} \label{PHmode1}
\int_{\Sigma} \delta N d^{3}x=0.
\end{equation}

\subsection{The perturbative constraints}

In this subsection, we will discuss the perturbative expressions of Gaussian constraint, diffeomorphism constraint and Hamiltonian constraint, respectively.

\subsubsection{Gaussian constraint}

In the connection dynamical formalism, the Gaussian constraint is given by
\begin{eqnarray} \label{GravityGaussC}
G[\Lambda] := \frac{1}{8\pi G\gamma}\int_{\Sigma}d^{3}x \Lambda^{i}G_{i}
=\frac{1}{8\pi G\gamma}\int_{\Sigma}d^{3}x \Lambda^{i}(\partial_{a}E^{a}_{i}+\epsilon_{ij}^{\ \ k}A^{j}_{a}E^{a}_{k}).
\end{eqnarray}
One can perturb it and get
\begin{eqnarray} \label{GravityGaussCPerturbation}
G[\Lambda]
=\frac{1}{8\pi G\gamma}\int_{\Sigma}d^{3}x \Lambda^{i}(\partial_{a}\delta E^{a}_{i}
+\epsilon_{ij}^{\ \ a}\bar{p}\delta A^{j}_{a}
+\epsilon_{ia}^{\ \ k}\bar{q}\delta E^{a}_{k})~.
\end{eqnarray}
Since internal gauge rotations of phase space functions $f$ are
parametrized by the smearing function $\Lambda^{i}$ in terms of $\delta_{\Lambda}f=\{f,G[\Lambda]\}$,
one can calculate the internal gauge rotations of perturbed basic variables
$\delta A^{i}_{a}$ and $\delta E^{a}_{i}$ as
\begin{eqnarray} \label{InternalGaugeRotations}
&&
\delta_{\Lambda}(\delta A^{i}_{a})
:=\{\delta A^{i}_{a},G[\Lambda]\}
=\bar{q}\Lambda^{l}\epsilon_{la}^{\ \ i}+\partial_{a}\Lambda^{l}
=\bar{q}\Lambda^{l}\epsilon_{la}^{\ \ i}~,
\nonumber\\
&&
\delta_{\Lambda}(\delta E^{a}_{i})
:=\{\delta E^{a}_{i},G[\Lambda]\}
=\bar{p}\Lambda^{l}\epsilon_{li}^{\ \ a}~.
\end{eqnarray}
In the final equality of the first equation in the above equations, we have used the fact that as a scalar, $\Lambda^{i}$ only has the homogeneous mode for vector perturbation. In order to have invariant basic perturbed variables under the internal gauge rotations, we ask the perturbed variables to be symmetrized. Therefore, the physical quantities depend only on the symmetrized perturbed basic variables, $\delta A^{\ i)}_{(a}$ and $\delta E^{\ a)}_{(i}$.

\subsubsection{Diffeomorphism constraint}

In general, up to Gaussian constraint, the diffeomorphism constraint of GR can be expressed as
\begin{eqnarray} \label{GravityDiffeomorphismC}
D_{G}[N^{a}] := \frac{1}{8\pi G\gamma}\int_{\Sigma}d^{3}x N^{c}(-s F^{k}_{cd} E^{d}_{k}).
\end{eqnarray}
For Euclidean GR, it reads
\begin{eqnarray} \label{EGravityDiffeomorphismC}
D_{EG}[N^{a}] &:=& \frac{1}{8\pi G}\int_{\Sigma}d^{3}x N^{c}(- F^{k}_{cd} E^{d}_{k})
\nonumber\\ \label{EGravityDiffeomorphismC1} &=& \frac{1}{8\pi G}\int_{\Sigma}d^{3}x N^{c}[(- \partial_{c}A^{k}_{d} + \partial_{d}A^{k}_{c})E^{d}_{k}
+A^{i}_{c}\partial_{a}E^{a}_{i}]
\end{eqnarray}
where the Gaussian constraint (\ref{GravityGaussC}) is used in the second equality.
Then, the perturbed diffeomorphism constraint can be expressed up to second order in perturbations as
\begin{eqnarray} \label{EGravityDiffeomorphismCPerturbation}
D_{EG}[N^{a}] = \frac{1}{8\pi G}\int_{\Sigma}d^{3}x \delta N^{c}
[\bar{p} (\partial_{k} \delta A^{k}_{c})
+ \bar{q} \delta^{k}_{c} (\partial_{d} \delta E^{d}_{k})].
\end{eqnarray}

\subsubsection{Hamiltonian constraint}

In the previous section, we have discussed Hamiltonian constraint with holonomy corrections. Now we will turn to discuss the perturbative Hamiltonian constraint in connection dynamical formalism.\footnote{Here we only give the expression of the pertrubative Hamiltonian density for vector modes, and its detailed derivation will be given in Appendix \ref{Appendix1}.}
Using the perturbed basic variables, we can expand the Euclidean
gravitational Hamiltonian density (\ref{GravityHamiltonianD}) up to the second order as
$\mathcal{H}_{EG}=\mathcal{H}_{EG}^{(0)}+ \mathcal{H}_{EG}^{(1)}+\mathcal{H}_{EG}^{(2)}$ with
\begin{eqnarray} \label{E0}
\mathcal{H}_{EG}^{(0)} &=& 6\bar{q}^{2}\sqrt{\bar p}~,
\nonumber\\ \label{E1} \mathcal{H}_{EG}^{(1)} &=&
2\sqrt{\bar{p}}\epsilon_{\ \ i}^{cd}\partial_{c}\delta A^{i}_{d}~,
\nonumber\\ \label{E2} \mathcal{H}_{EG}^{(2)} &=&
-\sqrt{\bar{p}} \delta A_c^j\delta A_d^k\delta^c_k\delta^d_j
+\frac{2\bar{q}}{\sqrt{\bar{p}}} \delta E^c_j\delta A_c^j
+\frac{\bar{q}^{2}}{2\bar{p}^{3/2}} \delta E^c_j\delta E^d_k\delta_c^k\delta_d^j
+ \frac{4}{\sqrt{\bar{p}}} \epsilon^{ck}_{\ \ i} \delta E^{d}_{k} \partial_{[c} \delta A^{i}_{d]}
~.
\end{eqnarray}
Similarly, the corrected Hamiltonian density (\ref{EuclideanHdwithHolonomy})
can be expressed up to the second order as
\begin{eqnarray} \label{h0}
\mathcal{H}_{EG}^{Q(0)} &=& \sqrt{\bar p}f^{i(0)}_{cd}\epsilon^{cd}_{\ \ i}~,
\nonumber\\ \label{h1} \mathcal{H}_{EG}^{Q(1)} &=&
\sqrt{\bar{p}}f^{i(1)}_{cd}\epsilon^{cd}_{\ \ i}
+ \frac{2}{\sqrt{\bar{p}}} f^{i(0)}_{cd} \epsilon^{ck}_{\ \ i} \delta E^{d}_{k}
~,
\nonumber\\ \label{h2} \mathcal{H}_{EG}^{Q(2)} &=&
\sqrt{\bar{p}} f^{i(2)}_{cd}\epsilon^{cd}_{\ \ i}
+\frac{2}{\sqrt{\bar{p}}} f^{i(1)}_{cd} \epsilon^{ck}_{\ \ i} \delta E^{d}_{k}
+\frac{f^{i(0)}_{cd}}{\bar{p}^{3/2}} \left( \epsilon_{\ \ i}^{jk} \delta E^{c}_{j} \delta E^{d}_{k}
+\frac{1}{4} \epsilon_{\ \ i}^{cd} \delta E^{a}_{j} \delta E^{b}_{k} \delta^{j}_{b} \delta^{k}_{a} \right)
~.
\end{eqnarray}
For simplicity, we denote $\mathcal{F}^{(0)}\equiv f^{i(0)}_{cd}\epsilon^{cd}_{\ \ i}$,
$\mathcal{F}^{(1)}\equiv f^{i(1)}_{cd}\epsilon^{cd}_{\ \ i}$ and $\mathcal{F}^{(2)}\equiv f^{i(2)}_{cd}\epsilon^{cd}_{\ \ i}$.
Then the above corrected Hamiltonian density can be reexpressed as
\begin{eqnarray} \label{h0F}
\mathcal{H}_{EG}^{Q(0)} &=& \sqrt{\bar p}\mathcal{F}^{(0)}~,
\nonumber\\ \label{h1} \mathcal{H}_{EG}^{Q(1)} &=&
\sqrt{\bar{p}}\mathcal{F}^{(1)}
+ \frac{2}{\sqrt{\bar{p}}} f^{i(0)}_{cd} \epsilon^{ck}_{\ \ i} \delta E^{d}_{k}
~,
\nonumber\\ \label{h2} \mathcal{H}_{EG}^{Q(2)} &=&
\sqrt{\bar{p}}\mathcal{F}^{(2)}
+ \frac{2}{\sqrt{\bar{p}}} f^{i(1)}_{cd} \epsilon^{ck}_{\ \ i} \delta E^{d}_{k}
+ \frac{1}{ 4 \bar{p}^{3/2}} \mathcal{F}^{(0)} \delta E^a_j \delta E^b_k \delta^j_b \delta^k_a
+ \frac{1}{\bar{p}^{3/2}} f^{i(0)}_{cd} \epsilon^{jk}_{\ \ i} \delta E^c_j \delta E^d_k
~.
\end{eqnarray}
It is easy to check that when $f^{i}_{cd}=F^{i}_{cd}$, the above corrected Hamiltonian constraint recovers Eq.(\ref{E0}).

\subsection{Constraint algebra}

Since the perturbed variables
do not have homogeneous modes as described in Eqs. (\ref{PHmode})
and (\ref{PHmode1}) and the boundary condition
that the integration over the boundary vanishes is required,
the integration $\int_{\Sigma} d^{3}x \bar{N} \mathcal{H}_{EG}^{(1)}$ and
$\int_{\Sigma} d^{3}x \delta N \mathcal{H}_{EG}^{(0)}$ vanishes.
Therefore, the explicit expression for the perturbed Hamiltonian constraint becomes
\begin{eqnarray}
\label{EuclideanPHC}
H_{EG}[\bar{N}]=\frac{1}{16\pi G}\int d^{3}x\bar{N}[\mathcal{H}_{EG}^{(0)}+\mathcal{H}_{EG}^{(2)}].
\end{eqnarray}
For the same reason, the expression for the corrected perturbed Hamiltonian constraint becomes
\begin{eqnarray}
\label{EuclideanCPHC1}
H_{EG}^{Q}[\bar{N}]
= \frac{1}{16\pi G}\int d^{3}x\bar{N}[\mathcal{H}_{EG}^{Q(0)}+\mathcal{H}_{EG}^{Q(2)}]~.
\end{eqnarray}
Since there is no lapse perturbations for the vector mode,
the Poisson bracket between the corrected Hamiltonian constraints,
$\{H_{EG}^{Q}[N_{1}],H_{EG}^{Q}[N_{2}]\}$, is trivial.
However, a nontrivial anomaly might occur in the Poisson bracket
between the corrected Hamiltonian constraint and the diffeomorphism constraint,
$\{H_{EG}^{Q}[N],D_{EG}[N^{a}]\}$.
In the following, we will discuss the conditions for an anomaly-free constraint algebra.

For simplicity, in this paper we will only consider the case that the holonomy-correction function $f^{i}_{cd}$
is a function of the connection variable $A^{m}_{a}$
and its first-order derivative, i.e., $f^{i}_{cd}\equiv f^{i}_{cd}(A,\partial A)$.
In this case, using the Taylor expansion, we can explicitly express the holonomy-correction function as
\begin{eqnarray} \label{HCFExpansionCaseI}
&&f^{i}_{cd}(A,\partial A,\epsilon)
\nonumber\\
&&= f^{i}_{cd}(\bar{A},\epsilon)
+\frac{\partial f^{i}_{cd}(A,\partial A,\epsilon)}{\partial A^{m}_{a}}|_{\bar{A}^{m}_{a}} \delta A^{m}_{a}
+\frac{\partial f^{i}_{cd}(A,\partial A,\epsilon)}{\partial (\partial_{e} A^{m}_{a})}|_{\bar{A}^{m}_{a}}
\partial_{e} \delta A^{m}_{a}
+\frac{1}{2}\frac{\partial^{2} f^{i}_{cd}(A,\partial A,\epsilon)}{\partial A^{m}_{a} \partial A^{n}_{b}}|_{\bar{A}^{m}_{a}} \delta A^{m}_{a} \delta A^{n}_{b}
\nonumber\\
&&
+\frac{\partial^{2} f^{i}_{cd}(A,\partial A,\epsilon)}{\partial A^{m}_{a} \partial (\partial_{e} A^{n}_{b})}|_{\bar{A}^{m}_{a}} \delta A^{m}_{a} \partial_{e}\delta A^{n}_{b}
+\frac{1}{2}\frac{\partial^{2} f^{i}_{cd}(A,\partial A,\epsilon)}{\partial (\partial_{e}A^{m}_{a}) \partial (\partial_{f} A^{n}_{b})}|_{\bar{A}^{m}_{a}} \partial_{e}\delta A^{m}_{a} \partial_{f}\delta A^{n}_{b}
+\ldots
\nonumber\\
&&=f^{i(0)}_{cd}(\bar{q},\epsilon)
+\mathcal{A}^{i(1)}_{cd}(\bar{q},\delta A,\epsilon)
+\mathcal{B}^{i(1)}_{cd}(\bar{q},\partial\delta A,\epsilon)
+\mathcal{A}^{i(2)}_{cd}(\bar{q},\delta A,\epsilon)
+\mathcal{C}^{i(2)}_{cd}(\bar{q},\delta A,\partial\delta A,\epsilon)
\nonumber\\
&&
+\mathcal{B}^{i(2)}_{cd}(\bar{q},\partial\delta A,\epsilon)
+\ldots
\end{eqnarray}
We also denote $f^{i(1)}_{cd}\equiv\mathcal{A}^{i(1)}_{cd}
+\mathcal{B}^{i(1)}_{cd}$ and $f^{i(2)}_{cd}\equiv\mathcal{A}^{i(2)}_{cd}+\mathcal{C}^{i(2)}_{cd}
+\mathcal{B}^{i(2)}_{cd}$. Therefore, the Poisson bracket between the corrected Hamiltonian
constraint and the diffeomorphism constraint can be calculated as
\begin{eqnarray} \label{PoissonBCHDCaseI}
&&\{H_{EG}^{Q}[N],D_{EG}[N^{a}]\}
\nonumber\\
&&
= \frac{1}{16\pi G}\int d^{3}x \delta N^{c}[-\frac{2}{3}\mathcal{F}^{(0)}\delta^{k}_{c}(\partial_{d}\delta E^{d}_{k})
-2 f^{i(0)}_{cd} \epsilon^{jk}_{\ \ i} \partial_{j} \delta E^{d}_{k}
+2 \bar{q} \frac{\partial f^{j(1)}_{bd}}{\partial(\delta A^{i}_{a})}
\epsilon^{bk}_{\ \ j}\delta^{i}_{c}\partial_{a}\delta E^{d}_{k}
\nonumber\\
&&
-2 \bar{q} \frac{\partial f^{j(1)}_{bd}}{\partial (\partial_{e}\delta A^{i}_{a})}\epsilon^{bk}_{\ \ j}\delta^{i}_{c}\partial_{a}\partial_{e}\delta E^{d}_{k}
+\frac{1}{3}\bar{p}\frac{\partial \mathcal{F}^{(0)}}{\partial \bar{q}}\partial_{k} \delta A^{k}_{c}
+\bar{q}\bar{p}\delta^{i}_{c}\partial_{a}\frac{\partial \mathcal{F}^{(2)}}{\partial (\delta A^{i}_{a})}
-\bar{q}\bar{p}\delta^{i}_{c}\partial_{a}\partial_{e}\frac{\partial \mathcal{F}^{(2)}}{\partial (\partial_{e}\delta A^{i}_{a})}
-2 \bar{p} \epsilon^{bj}_{\ \ i}\partial_{j}f^{i(1)}_{bc}]
\nonumber\\
&&= \frac{1}{16\pi G}\int d^{3}x \delta N^{c}[-\frac{2}{3}\mathcal{F}^{(0)}\delta^{k}_{c}(\partial_{d}\delta E^{d}_{k})
-2 f^{i(0)}_{cd} \epsilon^{jk}_{\ \ i} \partial_{j} \delta E^{d}_{k}
+2 \bar{q} \frac{\partial \mathcal{A}^{j(1)}_{bd}}{\partial(\delta A^{i}_{a})}
\epsilon^{bk}_{\ \ j}\delta^{i}_{c}\partial_{a}\delta E^{d}_{k}
\nonumber\\
&& \quad
-2 \bar{q} \frac{\partial \mathcal{B}^{j(1)}_{bd}}{\partial (\partial_{e}\delta A^{i}_{a})}\epsilon^{bk}_{\ \ j}\delta^{i}_{c}\partial_{a}\partial_{e}\delta E^{d}_{k}
+\frac{1}{3}\bar{p}\frac{\partial \mathcal{F}^{(0)}}{\partial \bar{q}}\partial_{k} \delta A^{k}_{c}
+\bar{q}\bar{p}\delta^{i}_{c}\partial_{a}\frac{\partial \mathcal{A}^{(2)}}{\partial (\delta A^{i}_{a})}
-2 \bar{p} \epsilon^{bj}_{\ \ i}\partial_{j}\mathcal{A}^{i(1)}_{bc}
+\bar{q}\bar{p}\delta^{i}_{c}\partial_{a}\frac{\partial \mathcal{C}^{(2)}}{\partial (\delta A^{i}_{a})}
\nonumber\\
&& \quad
-\bar{q}\bar{p}\delta^{i}_{c}\partial_{a}\partial_{e}\frac{\partial \mathcal{C}^{(2)}}{\partial (\partial_{e}\delta A^{i}_{a})}
-2 \bar{p} \epsilon^{bj}_{\ \ i}\partial_{j}\mathcal{B}^{i(1)}_{bc}
-\bar{q}\bar{p}\delta^{i}_{c}\partial_{a}\partial_{e}\frac{\partial \mathcal{B}^{(2)}}{\partial (\partial_{e}\delta A^{i}_{a})}]
\end{eqnarray}
To avoid anomaly, we require the Poisson bracket (\ref{PoissonBCHDCaseI}) to be closed. This means that the above Poisson bracket should be expressed as a linear combination of the Hamiltonian constraint and the diffeomorphism constraint or vanish. Since the holonomy-correction function $f^{i}_{cd}$ is in principle computable in the full theory,
the above requirement provides an important consistency check for LQG.
It may exclude certain forms of $f^{i}_{cd}$ or put some constraints on them.
Now a question immediately occurs: does there exist at all any nontrival form of $f^{i}_{cd}$
meeting the above requirement?

\subsection{The construction of $f^{i}_{cd}$}

We consider the following construction of the holonomy-corrected function:
\begin{eqnarray}\label{ficdconcrete}
f^{i}_{cd}&=&\sigma(\bar{q})\epsilon^{\ \ i}_{cd}
+\slashed{\sigma}(\bar{q})\epsilon^{\ \ i}_{cd}A^{j}_{a}\delta^{a}_{j}
+\mu(\bar{q}) A^{i}_{b}\epsilon^{\ \ b}_{cd}
+ \nu(\bar{q})(\epsilon^{\ \ i}_{md} A^{m}_{c} + \epsilon^{\ \ i}_{cn} A^{n}_{d})
\nonumber\\
&&
+\widetilde{\slashed{\sigma}}(\bar{q})\epsilon^{\ \ i}_{cd}(A^{j}_{a}\delta^{a}_{j})^{2}
+\slashed{\mu}(\bar{q}) A^{i}_{b}\epsilon^{\ \ b}_{cd}A^{j}_{a}\delta^{a}_{j}
+\slashed{\nu}(\bar{q})(\epsilon^{\ \ i}_{md} A^{m}_{c} + \epsilon^{\ \ i}_{cn} A^{n}_{d})A^{j}_{a}\delta^{a}_{j}
\nonumber\\
&&
+ \beta(\bar{q}) \epsilon^{\ \ i}_{mn}A^{m}_{c}A^{n}_{d}
+ \alpha(\bar{q})\partial_{[c} A^{i}_{d]}
.
\end{eqnarray}
Note that we only consider the terms up to second order in the holonomy-corrected function $f^{i}_{cd}$.
In addition, one can easily check that the corrected function $f^{i}_{cd}$ is antisymmetric, i.e., $f^{i}_{cd}=-f^{i}_{dc}$.
With a concrete holonomy-corrected function $f^{i}_{cd}$ at hand, we can calculate the Poisson bracket (\ref{PoissonBCHDCaseI}) between the corrected Hamiltonian constraint and the diffeomorphism constraint.
First, we have
\begin{eqnarray} \label{f1expand}
f^{i(0)}_{cd} &=& (\sigma + 3\slashed{\sigma} \bar{q} + \mu \bar{q} + 2 \nu \bar{q} + \beta\bar{q}^{2}
+9\widetilde{\slashed{\sigma}}\bar{q}^{2}
+3\slashed{\mu}\bar{q}^{2}+6\slashed{\nu}\bar{q}^{2}
)\epsilon^{\ \ i}_{cd},
\nonumber\\
\mathcal{A}^{i(1)}_{cd} &=&
(\nu+\beta\bar{q}+3\slashed{\nu}\bar{q})
(\epsilon^{\ \ i}_{md}\delta A^{m}_{c}+\epsilon^{\ \ i}_{cn}\delta A^{n}_{d})
+(\mu+3\slashed{\mu}\bar{q})\delta A^{i}_{b}\epsilon^{\ \ b}_{cd},
~~~
\nonumber\\
\mathcal{B}^{i(1)}_{cd} &=& \alpha \partial_{[c}\delta A^{i}_{d]}
,
\nonumber\\
\mathcal{A}^{i(2)}_{cd} &=&
\beta\epsilon^{\ \ i}_{mn}\delta A^{m}_{c} \delta A^{n}_{d}
,
\nonumber\\
\mathcal{B}^{i(2)}_{cd}&=&0,
\nonumber\\
\mathcal{C}^{i(2)}_{cd}&=&
0.
\end{eqnarray}
In the above equations, we have used the divergence-free property, i.e., $\delta^{b}_{j} \delta A^{j}_{b}=0$ and $\delta^{i}_{a} \delta E^{a}_{i}=0$, for vector mode. In addition, for convenience, we list some necessary relations: %
\begin{eqnarray} \label{ficdexpandNecessaryRelationVector}
\mathcal{F}^{(0)} &=& 6(\sigma + 3\slashed{\sigma} \bar{q} + \mu \bar{q} + 2 \nu \bar{q} + \beta\bar{q}^{2}
+9\widetilde{\slashed{\sigma}}\bar{q}^{2}
+3\slashed{\mu}\bar{q}^{2}+6\slashed{\nu}\bar{q}^{2}
),~~~~~f^{i(0)}_{cd}=\frac{\mathcal{F}^{(0)}}{6}\epsilon^{\ \ i}_{cd},
\nonumber\\
\mathcal{A}^{(1)} &=&
0,
~~~~~
\mathcal{B}^{(1)} = 0
,
~~~~~
\mathcal{A}^{(2)} =
-\beta \delta A^{m}_{c}\delta A^{n}_{d}\delta^{c}_{n}\delta^{d}_{m}
.
\end{eqnarray}
Substituting (\ref{f1expand}) and (\ref{ficdexpandNecessaryRelationVector}) into (\ref{PoissonBCHDCaseI}), we have
\begin{eqnarray} \label{PoissonBCHDFresult}
&&\{H_{EG}^{Q}[\bar{N}],D_{EG}[N^{a}]\}
\nonumber\\
&&
= \frac{1}{16\pi G}\int d^{3}x \delta N^{c}
[-2(\frac{\sigma}{\bar{q}} + 3\slashed{\sigma} + 2\mu + \nu
+9\bar{q}\widetilde{\slashed{\sigma}}
+6\bar{q}\slashed{\mu}
+3\bar{q}\slashed{\nu}
)\bar{q}\delta^{k}_{c}(\partial_{d}\delta E^{d}_{k})
\nonumber\\
&&
+2(\frac{\partial\sigma}{\partial \bar{q}}
+3\bar{q}\frac{\partial\slashed{\sigma}}{\partial \bar{q}}
+\bar{q} \frac{\partial\mu}{\partial \bar{q}}
+2\bar{q} \frac{\partial\nu}{\partial \bar{q}}
+3\bar{q}^{2}\frac{\partial\slashed{\mu}}{\partial \bar{q}}
+\bar{q}^{2}\frac{\partial\beta}{\partial \bar{q}}
+9\bar{q}^{2}\frac{\partial \widetilde{\slashed{\sigma}}}{\partial \bar{q}}
+6\bar{q}^{2}\frac{\partial \slashed{\nu}}{\partial \bar{q}}
\nonumber\\
&&
+3\slashed{\sigma}+2\mu+\nu+18\bar{q}\widetilde{\slashed{\sigma}}
+9\bar{q}\slashed{\mu}
+9\bar{q}\slashed{\nu}
)\bar{p}\partial_{k}\delta A^{k}_{c}.
\end{eqnarray}
If we impose the condition
\begin{eqnarray} \label{HDAnomalytermcancel1Vector}
&&
\frac{\partial\sigma}{\partial \bar{q}}
+3\bar{q}\frac{\partial\slashed{\sigma}}{\partial \bar{q}}
+\bar{q} \frac{\partial\mu}{\partial \bar{q}}
+2\bar{q} \frac{\partial\nu}{\partial \bar{q}}
+3\bar{q}^{2}\frac{\partial\slashed{\mu}}{\partial \bar{q}}
+\bar{q}^{2}\frac{\partial\beta}{\partial \bar{q}}
+9\bar{q}^{2}\frac{\partial \widetilde{\slashed{\sigma}}}{\partial \bar{q}}
+6\bar{q}^{2}\frac{\partial \slashed{\nu}}{\partial \bar{q}}
\nonumber\\
&&
=-\frac{\sigma}{\bar{q}} - 6\slashed{\sigma} - 4\mu - 2\nu
-27\bar{q}\widetilde{\slashed{\sigma}}
-15\bar{q}\slashed{\mu}
-12\bar{q}\slashed{\nu},
\end{eqnarray}
we will obtain a closed Poisson bracket as
\begin{eqnarray}
\{H_{EG}^{Q}[N],D_{EG}[N^{a}]\}
= -(\frac{\sigma}{\bar{q}} + 3\slashed{\sigma} + 2\mu + \nu
+9\bar{q}\widetilde{\slashed{\sigma}}
+6\bar{q}\slashed{\mu}
+3\bar{q}\slashed{\nu}
)D_{EG}[N^{a}]
.
\end{eqnarray}
Now, we have obtained a closed constraint algebra between the holonomy-corrected Hamiltonian constraint and the diffeomorphism constraint, which implies that in the perturbation framework, we can have nontrivial holonomy corrections from full theory while the constraint algebra is closed.

\section{Concluding remarks}

In order to consider the perturbations
in a framework containing holonomy correction from full LQG, we propose an effective holonomy-corrected Hamiltonian of full theory in the Euclidean GR. We have derived the Poisson bracket between the corrected Hamiltonian constraint and the diffeomorphism constraint for vector modes. We have also found a specific form of the holonomy-correction function $f^{i}_{cd}$, which satisfies that the constraint algebra is closed. As the first step, our result confirms, in the perturbation framework, the possibility of nontrivial holonomy corrections from full theory while the anomaly-free constraint algebra is preserved. This is a valuable and positive hint to the consistency of perturbative LQG. However, whether such form of holomomy correction can be strictly derived from the full LQG is still an open issue.

There are several directions for future work. It is desirable and important to calculate the perturbative constraint algebra for scalar modes in this framework\cite{HolonomyScalarmodeJPWu}. As one expected, further constraints on the holonomy correction function $f^{i}_{cd}$ would be found. It is also interesting and important to derive the corresponding cosmological perturbation equations.
The extension of our setup in this paper to the Lorentzian case would be more interesting and valuable, as it is the case of most interest in our Universe. In this case, the construction of the effective holonomy-corrected Hamiltonian of full LQG and the calculation of constraint algebra would be more complicated. We thus leave them for future study.

\section*{Acknowledgements}

We would like to thank Martin Bojowald for helpful discussions and suggestions. This work is supported in part by NSFC (Grants No. 10975017 and No. 11235003) and the Fundamental Research Funds for the Central Universities.

\begin{appendix}

\section{The perturbed Hamiltonian constraint}\label{Appendix1}

In this appendix, we derive the perturbed Hamiltonian constraint up to the second-order term of
the phase space variables $(A^{j}_{b},E^{a}_{i})$.
To this aim, we need the expansion of $(det E)^{-\frac{1}{2}}$ up to the second order.
Since $det E = \frac{1}{6}\epsilon_{abc}\epsilon^{ijk}E^{a}_{i}E^{b}_{j}E^{c}_{k}$, we have
\begin{eqnarray} \label{detEexpansion}
&&(det E)^{-\frac{1}{2}}
\nonumber\\
&&
=(det E)^{-\frac{1}{2}}\mid_{\bar{E}^{a}_{i}}
+\frac{\partial (det E)^{-\frac{1}{2}}}{\partial E^{a}_{i}}\mid_{\bar{E}^{a}_{i}}\delta E^{a}_{i}
+\frac{1}{2}\frac{\partial^{2} (det E)^{-\frac{1}{2}}}{\partial E^{a}_{i}\partial E^{b}_{j}}\mid_{\bar{E}^{a}_{i}}\delta E^{a}_{i}\delta E^{b}_{j}+\ldots
\nonumber\\
&&
=\bar{p}^{-\frac{3}{2}}\left[1-\frac{1}{2\bar{p}}\delta^{i}_{a}\delta E^{a}_{i}
+\frac{1}{8\bar{p}^{2}}(\delta^{i}_{a}\delta E^{a}_{i})^{2}
+\frac{1}{4\bar{p}^{2}}\delta E^{a}_{i}\delta E^{b}_{j}\delta^{i}_{b}\delta^{j}_{a}
+\ldots \right]
~.
\end{eqnarray}
By Eq.(\ref{detEexpansion}), one can expand the expression of
$\epsilon^{jk}_{\ \ i}\frac{E^{c}_{j}E^{d}_{k}}{\sqrt{|det E|}}$
up to the second order as
\begin{eqnarray} \label{EEdetEexpansion}
\left(\epsilon^{jk}_{\ \ i}\frac{E^{c}_{j}E^{d}_{k}}{\sqrt{|det E|}}\right)^{(0)}&=&\sqrt{\bar{p}}\epsilon^{cd}_{\ \ i}
\nonumber\\
\left(\epsilon^{jk}_{\ \ i}\frac{E^{c}_{j}E^{d}_{k}}{\sqrt{|det E|}}\right)^{(1)}
&=&-\frac{1}{2\sqrt{\bar{p}}}\epsilon^{cd}_{\ \ i}\delta^{j}_{a}\delta E^{a}_{j}
+\frac{1}{\sqrt{\bar{p}}}\epsilon^{ck}_{\ \ i}\delta E^{d}_{k}
+\frac{1}{\sqrt{\bar{p}}}\epsilon^{jd}_{\ \ i}\delta E^{c}_{j}
\nonumber\\
\left(\epsilon^{jk}_{\ \ i}\frac{E^{c}_{j}E^{d}_{k}}{\sqrt{|det E|}}\right)^{(2)}
&=&\frac{1}{8\bar{p}^{\frac{3}{2}}}\epsilon^{cd}_{\ \ i}(\delta^{j}_{a}\delta E^{a}_{j})^{2}
+\frac{1}{4\bar{p}^{\frac{3}{2}}}\epsilon^{cd}_{\ \ i}\delta E^{a}_{k}\delta E^{b}_{j}\delta^{k}_{b}\delta^{j}_{a}
-\frac{1}{2\bar{p}^{\frac{3}{2}}}\epsilon^{ck}_{\ \ i}\delta E^{d}_{k}\delta E^{a}_{j}\delta^{j}_{a}
\nonumber\\
&& \quad
-\frac{1}{2\bar{p}^{\frac{3}{2}}}\epsilon^{kd}_{\ \ i}\delta E^{c}_{k}\delta E^{a}_{j}\delta^{j}_{a}
+\frac{1}{\bar{p}^{\frac{3}{2}}}\epsilon^{jk}_{\ \ i}\delta E^{c}_{j}\delta E^{d}_{k}
~.
\end{eqnarray}
At the same time, expanding the curvature $F^{i}_{cd}$ up to the second order, we have
\begin{eqnarray} \label{Fexpansion}
&&F^{i(0)}_{cd}=\bar{q}^{2}\epsilon^{i}_{\ cd}
\nonumber\\
&&
F^{i(1)}_{cd}=2\partial_{[c}\delta A_{d]}^{i}
+\bar{q}\epsilon^{i}_{\ md}\delta A^{m}_{c}
+\bar{q}\epsilon^{i}_{\ cn}\delta A^{n}_{d}
\nonumber\\
&&
F^{i(2)}_{cd}=\epsilon^{i}_{\ mn}\delta A^{m}_{c}\delta A^{n}_{d}
~.
\end{eqnarray}
By using the above equations (\ref{EEdetEexpansion}) and (\ref{Fexpansion}),
one can obtain the expansions of the Hamiltonian density up to the second order as
\begin{eqnarray} \label{HamiltonianExpansion}
\mathcal{H}_{EG}^{(0)} &=& 6\bar{q}^{2}\sqrt{\bar p}~,
\nonumber\\
\mathcal{H}_{EG}^{(1)} &=&
4 \bar{q}\sqrt{\bar{p}} \delta^c_j\delta A_c^j
+\frac{\bar{q}^{2}}{\sqrt{\bar{p}}} \delta_c^j\delta E^c_j
+2\sqrt{\bar{p}}\epsilon_{\ \ i}^{cd}\partial_{c}\delta A^{i}_{d}~,
\nonumber\\
\mathcal{H}_{EG}^{(2)} &=&
-\sqrt{\bar{p}} \delta A_c^j\delta A_d^k\delta^c_k\delta^d_j
+\sqrt{\bar{p}} (\delta A_c^j\delta^c_j)^2
+\frac{2\bar{q}}{\sqrt{\bar{p}}} \delta E^c_j\delta A_c^j
+\frac{\bar{q}^{2}}{2\bar{p}^{3/2}} \delta E^c_j\delta E^d_k\delta_c^k\delta_d^j
\nonumber\\
&& \quad
-\frac{\bar{q}^{2}}{4\bar{p}^{3/2}}(\delta E^c_j\delta_c^j)^2
+\frac{1}{\sqrt{\bar{p}}}\left( 4 \epsilon^{ck}_{\ \ i} \delta E^{d}_{k} - \epsilon^{cd}_{\ \ i} \delta E^{a}_{j} \delta^{j}_{a} \right) \partial_{[c} \delta A^{i}_{d]}
 ~.
\end{eqnarray}
Since $\delta^{b}_{j} \delta A^{j}_{b}=0$ and $\delta^{i}_{a} \delta E^{a}_{i}=0$ for vector modes,
the above expansions reduce to Eq.(\ref{E0}).
Now, we consider the corrected Hamiltonian.
First, up to the second order, one can expand the holonomy-correction function $f^{i}_{cd}$ as
\begin{eqnarray} \label{HCFExpansionT}
&&f^{i}_{cd}(A,\partial A,\cdots,\partial^{n} A,\epsilon)
\nonumber\\
&&
=f^{i(0)}_{cd}(\bar{q},\epsilon)
+f^{i(1)}_{cd}(\bar{q},\delta A, \partial \delta A,\cdots,\partial^{n} \delta A,\epsilon)
+f^{i(2)}_{cd}(\bar{q},\delta A, \partial \delta A,\cdots,\partial^{n} \delta A,\epsilon)
+\ldots.
\end{eqnarray}
By the equations (\ref{EEdetEexpansion}) and (\ref{HCFExpansionT}),
the holonomy-correction Hamiltonian density $\mathcal{H}^{Q}_{EG}$ (\ref{EuclideanHdwithHolonomy})
can be expressed up to the second order as
\begin{eqnarray} \label{QHamiltonianExpansion}
\mathcal{H}_{EG}^{Q(0)} &=& \sqrt{\bar p}f^{i(0)}_{cd}\epsilon^{cd}_{\ \ i}~,
\nonumber\\ \mathcal{H}_{EG}^{Q(1)} &=&
\sqrt{\bar{p}}f^{i(1)}_{cd}\epsilon^{cd}_{\ \ i}
+\frac{f^{i(0)}_{cd}}{\sqrt{\bar{p}}} \left( 2 \epsilon^{ck}_{\ \ i} \delta E^{d}_{k}
- \frac{1}{2} \epsilon^{cd}_{\ \ i} \delta E^{a}_{j} \delta^{j}_{a} \right)
~,
\nonumber\\ \mathcal{H}_{EG}^{Q(2)} &=&
\sqrt{\bar{p}} f^{i(2)}_{cd}\epsilon^{cd}_{\ \ i}
+\frac{f^{i(1)}_{cd}}{\sqrt{\bar{p}}} \left( 2 \epsilon^{ck}_{\ \ i} \delta E^{d}_{k}
- \frac{1}{2} \epsilon^{cd}_{\ \ i} \delta E^{a}_{j} \delta^{j}_{a} \right)
\nonumber\\
&& \quad
+\frac{f^{i(0)}_{cd}}{\bar{p}^{3/2}} \left[ \epsilon_{\ \ i}^{jk} \delta E^{c}_{j} \delta E^{d}_{k}
-\epsilon_{\ \ i}^{ck} \delta E^{d}_{k} \delta E^{a}_{j} \delta^{j}_{a}
+\frac{1}{8} \epsilon_{\ \ i}^{cd} (\delta E^{a}_{j} \delta^{j}_{a})^{2}
+\frac{1}{4} \epsilon_{\ \ i}^{cd} \delta E^{a}_{j} \delta E^{b}_{k} \delta^{j}_{b} \delta^{k}_{a} \right]
.
\end{eqnarray}
Furthermore, if we denote $\mathcal{F}^{(0)}=f^{i(0)}_{cd}\epsilon^{cd}_{\ \ i}$,
$\mathcal{F}^{(1)}=f^{i(1)}_{cd}\epsilon^{cd}_{\ \ i}$ and $\mathcal{F}^{(2)}=f^{i(2)}_{cd}\epsilon^{cd}_{\ \ i}$,
the above corrected Hamiltonian constraint can be reexpressed as
\begin{eqnarray} \label{QHamiltonianExpansionF}
\mathcal{H}_{EG}^{Q(0)} &=& \sqrt{\bar p}\mathcal{F}^{(0)}~,
\nonumber\\ \mathcal{H}_{EG}^{Q(1)} &=&
\sqrt{\bar{p}}\mathcal{F}^{(1)}
-\frac{1}{ 2 \sqrt{\bar{p}}} \mathcal{F}^{(0)} \delta E^a_j \delta^j_a
+ \frac{2}{\sqrt{\bar{p}}} f^{i(0)}_{cd} \epsilon^{ck}_{\ \ i} \delta E^{d}_{k}
~,
\nonumber\\ \mathcal{H}_{EG}^{Q(2)} &=&
\sqrt{\bar{p}}\mathcal{F}^{(2)}
-\frac{1}{ 2 \sqrt{\bar{p}}} \mathcal{F}^{(1)} \delta E^a_j \delta^j_a
+ \frac{2}{\sqrt{\bar{p}}} f^{i(1)}_{cd} \epsilon^{ck}_{\ \ i} \delta E^{d}_{k}
+ \frac{1}{ 8 \bar{p}^{3/2}} \mathcal{F}^{(0)} (\delta E^a_j \delta^j_a)^{2}
\nonumber\\
&& \quad
+ \frac{1}{ 4 \bar{p}^{3/2}} \mathcal{F}^{(0)} \delta E^a_j \delta E^b_k \delta^j_b \delta^k_a
- \frac{1}{\bar{p}^{3/2}} f^{i(0)}_{cd} \epsilon^{ck}_{\ \ i} \delta E^d_k \delta E^a_j \delta^j_a
+ \frac{1}{\bar{p}^{3/2}} f^{i(0)}_{cd} \epsilon^{jk}_{\ \ i} \delta E^c_j \delta E^d_k
~,
\end{eqnarray}
Also, for vector modes, Eqs. (\ref{QHamiltonianExpansion}) and (\ref{QHamiltonianExpansionF}) will reduce to the expressions (\ref{h0}) and (\ref{h0F}) respectively.

\end{appendix}

\end{document}